\documentclass[prc,twocolumn,letterpaper,showpacs,amsmath,amssymb,floatfix,nofootinbib]{revtex4-1}
\usepackage{graphicx,bm,url,braket,xcolor,xspace}

\newcommand{\bb}{$\beta\beta$\xspace}
\newcommand{\bbz}{$0\nu\beta\beta$\xspace}

\begin{document}

\title{Nonperturbative renormalization of the neutrinoless double-beta
operator in $p$-shell nuclei}

\author{Deepshikha Shukla}
\author{Jonathan Engel}

\affiliation{Dept.\ of Physics and Astronomy, University of North Carolina,
Chapel Hill, NC, 27516-3255, USA}

\author{Petr Navratil}
\affiliation{TRIUMF, 4004 Wesbrook Mall, Vancouver, British Columbia, V6T 2A3
Canada}
\affiliation{Lawrence Livermore National
Laboratory, P.O. Box 808, L-414, Livermore, CA 94551, USA}

\begin{abstract}
We use Lee-Suzuki mappings and related techniques to construct effective
two-body $p$-shell interactions and neutrinoless double-beta operators that
exactly reproduce the results of large no-core-shell-model calculations of
double-beta decay in nuclei with mass number $A=6$.  We then apply the
effective operators to the decay of nuclei with $A=7$, 8, and 10, again
comparing with no-core calculations in much larger spaces.  The results with
the effective two-body operators are generally good.  In some cases, however,
they differ non-negligibly from the full no-core results, suggesting that
three-body corrections to the decay operator in heavier nuclei may be
important.  An application of our procedure and related ideas to $fp$-shell
nuclei such as $^{76}$Ge should be feasible within coupled-cluster theory.

\end{abstract}

\pacs{}
\keywords{}

\maketitle

\section{\label{s:introduction}Introduction}

Particle physicists hope to learn about the overall neutrino-mass scale by
observing neutrinoless double-beta (\bbz) decay \cite{avi08}.  To extract a
mass from a lifetime, however, one must know the value of the nuclear matrix
element that governs the decay.  For that reason, theorists have worked hard
over the last 20 years to better calculate the matrix elements.  

One of the best frameworks for the job at present is the nuclear shell model.
Good calculations (e.g.\ Refs.\ \cite{men08,horoi10}) use model spaces of
dimension $10^7$ or larger by including the full valence shell.  Even these
calculations, however, omit most of the relevant many-particle Hilbert space by
requiring that most particles remain frozen in an inert core and prohibiting
even active particles from sampling levels above the valence shell.  These
approximations induce error that can in principle be accounted for through the
use of an effective Hamiltonian and decay operator.  The literature contains a
number of schemes for constructing effective operators \cite{ell77,hjo95}.  In
practice, however, such techniques are generally restricted to the effective
Hamiltonian, and even there the application is through a perturbative scheme
whose weaknesses often must be supplemented by fitting to spectra.  The decay
operator is usually not corrected at all, except at short distances (and
sometimes through an overall phenomenological multiplication factor).  How much
are calculated matrix-element corrupted as a result?  How might one do better?

To begin to answer these questions, we look at the \bbz matrix element in
nuclei with mass number $A$ between 6 and 10.  Such nuclei, of course, do not
undergo \bb decay, but one can calculate the matrix elements nonetheless.
Moreover, in these nuclei we can carry out fairly complete no-core shell-model
(NCSM) \cite{nav09} calculations and map their results onto valence-shell
($p$-shell) calculations to construct effective operators that reproduce the
full matrix elements exactly.  We can then bypass perturbation theory, which is
often unreliable (and was applied inconclusively to \bb decay in Ref.\
\cite{eng09}), and test nonperturbative approximations to the full effective
operator.  Reference \cite{lis08,lis09} carried out this program for
charge-conserving electromagnetic transition operators, the leading pieces of
which are one-body.  Here, the lowest-order effective decay operator acts on
two bodies, and we examine the restriction to this leading term.  We define
effective operators that reproduce the exact matrix element for the artificial
decay of states in $^6$He to those in $^6$Be.  Are these operators
significantly different from their bare counterparts?  Can they also reproduce
\bbz matrix elements in heavier nuclei, or are three- and higher-body effective
operators necessary as well?  The answers will provide a good idea of how much
work awaits us in the heavier nuclei that actually undergo \bb decay.

In Sec.\ \ref{s:methods} below, we describe the concepts and methods we
employ.  Section \ref{s:results} presents our results, and Sec.\
\ref{s:discussion} discusses their implications for matrix elements in the
heavier nuclei that are used in \bb experiments.

\section{\label{s:methods}Methods}

In the closure approximation (which is good for neutrinoless decay) and the
usual assumption that the nuclear weak current is adequately represented by a
one-body operator, the \bbz matrix element is a sum of three terms:
\begin{align}
\label{eq:me}
\mathcal{M}_{fi} \equiv \bra{f} \sum_{ab} M_{ab}^{GT} + M_{ab}^F + M_{ab}^T \ket{i}\,,
\end{align}
the last of which is a very small tensor piece \cite{sim99} that will be
ignored here.  The other two $M$'s are given by \cite{sim99,sim08}
\begin{align}
\label{eq:gtf}
M_{ab}^{GT}&= H_{GT}(r_{ab}) \, \boldsymbol{\sigma}_a \cdot \boldsymbol{\sigma}_b\\ 
M_{ab}^{F}&= H_{F}(r_{ab}) \,, \nonumber
\end{align}
with the labels $a$ and $b$ indicating nucleons both here and in Eq.\
(\ref{eq:me}), $r_{ab}$ representing internucleon distance, and the ``neutrino
potentials'' $H$ defined by
\begin{align}
\label{eq:bbpots}
H_{K}(r)  &= \frac{2R}{\pi r} \int_0^{\infty} \frac{h_K(q) \sin{qr}}
{q + \bar{\omega}} \, dq \,, \quad K=GT,F \,.
\end{align}
The $h_K(q)$ in Eq.\ (\ref{eq:bbpots}) contain the vector and axial-vector
coupling constants, form factors that account for the finite size of the
nucleon, and the effects of forbidden currents (weak magnetism and the induced
pseudoscalar term).  The quantity $\bar{\omega}$ is an average
intermediate-nucleus excitation energy to which the $H_K$ are not very
sensitive.  The authors of Ref.\ \cite{men11} recently applied chiral
effective-field theory to derive two-body corrections to the weak current and
thus three-body corrections to the operators in Eq.\ (\ref{eq:gtf}).  From our
point of view, these corrections modify the bare \bbz operator and are subject
to the same nuclear-structure renormalization that we apply to the two-body
operators in Eq.\ (\ref{eq:gtf}).  We neglect the chiral corrections here to
keep matters simple.

The matrix element $\mathcal{M}_{fi}$ is often small because of cancellations
among contributions at different internucleon distances $r_{ab}$. Instead of
looking only at the matrix element, therefore, we also examine the internucleon
matrix-element distribution $C(r)$, defined e.g.\ in Ref.\ \cite{engel11}, so
that
\begin{align}
\label{eq:dist}
\int_0^\infty C(r) \, dr = \mathcal{M}_{fi} \,.
\end{align}

Our starting point for calculating matrix elements is the NCSM.  We use
different starting interactions --- the CD Bonn potential \cite{cdbonn} and the
N$^3$LO chiral effective-field-theory interaction \cite{idahon3lo} ---
and model spaces that allow between six and ten $\hbar\omega$ of excitation
energy outside the $p$ shell (i.e. the NCSM parameter
$\mathrm{N}_{\mathrm{max}}$ is between 6 and 10).  We first apply standard
Lee-Suzuki techniques \cite{oku54, suz80a, nav98} to the Bonn potential
and the Similarity Renormalization Group (SRG) \cite{bog06, bog07} to the
chiral potential to construct interactions appropriate for those model
spaces.  In principle the double-beta decay operator should be treated in the
same way.  Preliminary studies \cite{stetcu09} show, however, that the
renormalization is slight and confined to short distances as expected, and
instead of carrying it out here we simulate short-range effects through an
effective Jastrow function from Ref.\ \cite{sim09}.

Many of the isotopes we discuss are very weakly bound or unbound in reality,
and our representation of them in the oscillator-based NCSM distorts their
structure.  For our purposes, however, the poor representation is not
important; we want to examine the effect of moving to a much smaller model
space, and take the large-space calculations to be the ``exact'' results we
want to reproduce.

Our small model space consists of all but four particles residing anywhere in
the $0p$ shell and the rest forming an inert $0s$-shell core.  As in Refs.\
\cite{lis08,lis09}, we first equate the effective neutron $p_{3/2}$ and
$p_{1/2}$ single-particle energies to the two lowest-energy eigenvalues
produced by the full calculation in $^5$He, and the effective proton energies
to the corresponding eigenvalues in $^5$Li.  Then in the $A=6$ nuclei we use
the Lee-Suzuki procedure to map the two lowest $J^\pi=0^+$ states, the lowest
$1^+$ state, and the two lowest 2$^+$ states (all with $T=1$) onto
corresponding orthogonal $p$-shell states.  In doing so we have assumed isospin
conservation in our small-space calculation; breaking isospin would require
only the additional straightforward step of carrying out separate calculations
in He and Be.  

The Lee-Suzuki mapping, which comes as close as possible to making the
$p$-shell energy eigenstates the projections of the corresponding full-space
states without spoiling orthogonality, proceeds as follows.  We let $P$ project
onto the $d$-dimensional small space, let $Q \equiv 1-P$, and denote by
$\ket{p}$, $\ket{p'}$, $\ket{p_1}$, etc., states for that are contained
entirely in the small space (with an analogous convention defining $\ket{q}$,
$\ket{q'}$, etc.).  The $d$ orthogonal small-space states $\ket{\tilde{k}}$
corresponding to $d$ selected full-space eigenstates $\ket{k}$ are defined by
\begin{equation}
\label{eq:mapping}
\ket{\tilde{k}} \equiv M^{-\frac{1}{2}} \left( P + \omega^\dag \right)
\ket{k}\,,
\end{equation}
with
\begin{equation}
\label{eq:omega}
\bra{q} \omega \ket{p} = \sum_{k=0}^{d-1}
\braket{q| k} \braket{\underbar{k} | p} \,.
\end{equation}
and
\begin{equation}
M = P+\omega^\dag \omega = P \left(1+\omega^\dag \omega \right) 
P\,.
\end{equation}
In Eq.\ (\ref{eq:omega}) the $\braket{\underbar{k} | p}$ are defined as the
elements of the inverse of the matrix with the $d^2$ elements $\braket{p|k}$.
With these definitions the effective operator $O_{\textrm{eff}}$ in the
dimension-$d$ small space that gives the same matrix elements as (the $T_z=0$
analogue of) the decay operator $O$ in the full space, is
\begin{equation}
O_{\textrm{eff}} = M^{-\frac{1}{2}} ( P+\omega^\dag ) O (P+\omega) 
M^{-\frac{1}{2}} \,.
\end{equation}
The matrix elements of this operator can be written without reference to any
vectors $\ket{q}$ or any of the eigenstates $\ket{k}$ beyond the $d$ that are
mapped, as
\begin{align}
\label{eq:Ome}
\bra{p} O_{\textrm{eff}} \ket{p'} &= \\
& \hspace{-1.5cm}  \sum_{p_1,p_2,k,k'=0}^{d-1}  
\bra{p} M^{-\frac{1}{2}}
\ket{p_1} \braket{p_1 | \underbar{k}} \bra{k} O \ket{k'} \braket{\underbar{k}' |
p_2} \bra{p_2}M^{-\frac{1}{2}} \ket{p'} \,, \nonumber 
\end{align}
where the elements of $M$ can be written in the same fashion as
\begin{equation}
\label{eq:Mme}
\bra{p} M \ket{p'} = \sum_{k=0}^{d-1} \braket{p|\underbar{k}}
\braket{\underbar{k} | p'} \,.
\end{equation}
We then simply use isospin algebra to obtain the matrix elements of the real
$T_z=2$ effective decay operator, for which $p$ represents two protons and $p'$
two neutrons.  To get the effective interaction we carry out a similar
procedure but also include the $T=0$ states in $^6$Li to obtain a complete set
of matrix elements.  

\begin{figure}[t]
\includegraphics[width=\columnwidth]{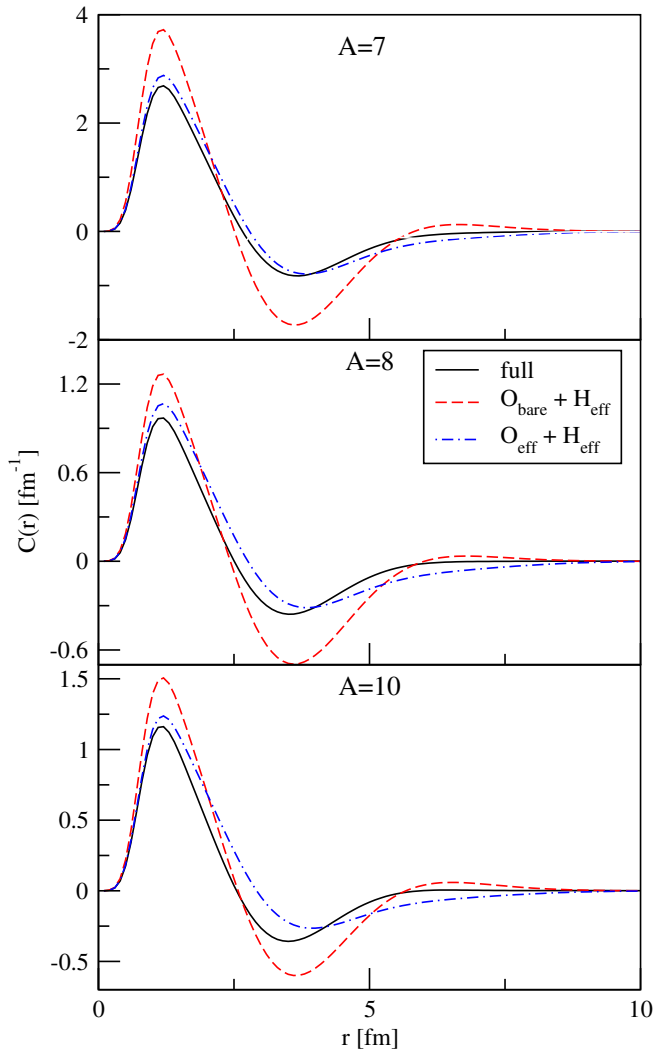}
\caption{\label{fig:srgn6} (Color online) The curves $C(r)$, the integral of
which gives the matrix element for neutrinoless double-beta decay.  The solid
(black) lines are the results of the full ($\mathrm{N}_{\mathrm{max}}=6$)
calculations with the SRG-evolved N$^3$LO potential, the dashed (red) lines are
the results of the $p$-shell calculation with the effective two-body
Hamiltonian and the bare decay operator, and the dot-dashed (blue) lines are
the results with the effective Hamiltonian and the effective decay operator.
The top panel is for the decay $^7$He $\rightarrow {}^7$Be, the middle panel
for $^8$He $\rightarrow {}^ 8$Be, and the bottom panel for $^{10}$He
$\rightarrow {}^{10}$Be.}
\end{figure}

The Lee-Suzuki mapping actually is only one of many that are possible.  As
noted above, the Lee-Suzuki procedure makes the small-space eigenvectors as
close as possible to projections of the full eigenvectors without sacrificing
orthogonality.  In other words, it constructs the orthonormal set
$\{\ket{\tilde{k}}\}$ that minimizes the quantity \cite{kvaal}
\begin{equation}
\label{eq:maxorthog}
\sum_{k=0}^{d-1} \left( \bra{k} - \bra{\tilde{k}}  \right)
\left( \ket{k} - \ket{\tilde{k}}  \right) \,.
\end{equation}
This prescription seems particularly appropriate for a comprehensive
description of the spectrum, but in double-beta decay we are not equally
interested in all states.  One alternative, known as the Contractor
Renormalization (CORE) mapping \cite{mue02} is to make the small-space ground
state $\ket{\tilde{0}}$ proportional $P \ket{0}$, the projection of the full
ground state, and then construct the other $\ket{\tilde{k}}$ from the set $P
\ket{k}$ through Graham-Schmidt orthogonalization.  But in fact any unitary
transformation of the $\ket{\tilde{k}}$'s generated by the Lee-Suzuki procedure
defines a valid mapping.  In our case, we can generate an arbitrary
time-reversal-preserving transformation by rotating the two small-space 0$^+$
states in $A=6$ by an arbitrary angle $\alpha$ and the two 2$^+$ states by
another angle $\beta$. We will try to see whether there are values of these
angles that are particularly suited for double-beta decay.

\section{\label{s:results}Results}

We now test the performance of our Lee-Suzuki effective operator in heavier
nuclei.  Figure \ref{fig:srgn6} presents our results for the decays
$^{7,8,10}$He$\rightarrow ^{7,8,10}$Be when we use the SRG-evolved chiral
N$^3$LO (\cite{idahon3lo}) interaction in a 6$\hbar\omega$ full space.  The
black (solid) curves in each of the panels denote the full
$\mathrm{N}_{\mathrm{max}}=6$ $0\nu\beta\beta$ distributions $C(r)$.  These
curves are what the effective operators are supposed to reproduce.  The red
(dashed) curves denote the result obtained with the bare $0\nu\beta\beta$
operator in the $p$ shell, with wave functions produced by the effective
$p$-shell interaction, which in turn comes from the Lee-Suzuki procedure for
$A=5$ and 6 discussed above.  The blue (dot-dashed) curves are the results with
the effective operator, used in conjunction with the wave functions from the
same effective interaction. 

\begin{table}[b]
\caption{Matix elements $\mathcal{M}_{fi}$ produced by the distributions $C(r)$
in Fig.~\ref{fig:srgn6}.}
\begin{tabular}{lccc}
\label{tab:srgn6}
& 7 & 8 & 10 \\
\hline
full & 1.76 & 0.48 & 0.79 \\
bare &  1.49 & 0.18 & 0.91 \\
effective & 1.90 & 0.59 & 1.23
\end{tabular}
\end{table}

The use of the effective decay operator clearly improves the agreement between
the $p$-shell $C(r)$ and the full one in all three panels.  One problem,
however, is that $C(r)$ is not itself measurable; its integral is what we
want.  And it turns out that oscillations can make apparent poor agreement
between curves much better in the integral, and good agreement worse.  Table
\ref{tab:srgn6} compares the matrix elements themselves for the three decays
represented by the figure.

The effective operator produces a clear improvement in the integrated matrix
element in A$=7$ and (particularly) 8, but by A$=10$ the bare operator does
pretty well and the effective operator not as well.  The reason is apparent
from the bottom panel of Fig.\ \ref{fig:srgn6}: the effective-operator curve,
while a better approximation than the bare curve, is not as good when
integrated because because it is above the full curve until about $r \sim 4$
fm.  The bare curve strays from the full curve at both the peak and dip but in
opposite directions; it thus supplies a good approximation when integrated.

\begin{figure}[t]
\includegraphics[width=\columnwidth]{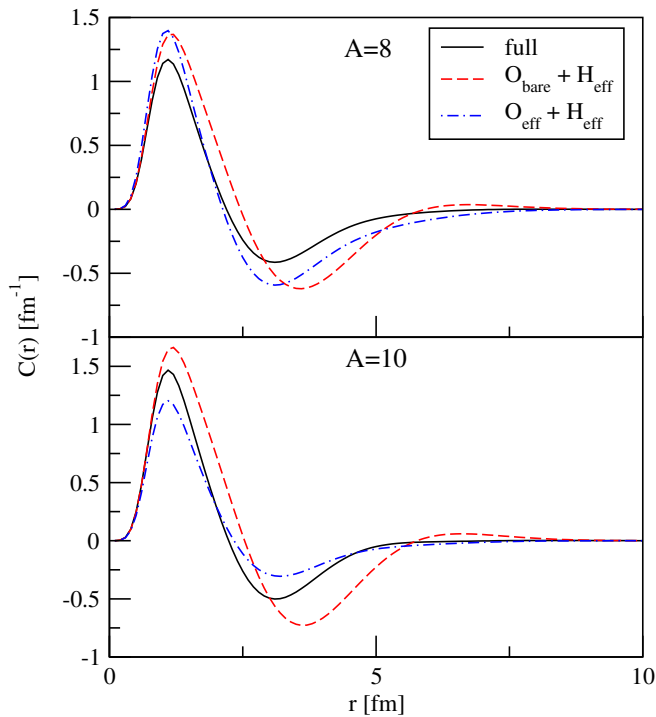}
\caption{\label{fig:cdb} (Color online) Same as Fig.\ \ref{fig:srgn6} but for
for $^8$He $\rightarrow {}^ 8$Be and $^{10}$He $\rightarrow {}^{10}$Be only and
with the CD-Bonn interaction conditioned for $\mathrm{N}_{\mathrm{max}}=8$ in
$A=8$ and $\mathrm{N}_{\mathrm{max}}=6$ in $A=10$. }
\end{figure}

Is this behavior a fluke?  Does it depend on the shell-model interaction or the
size of the full model space?  To address the questions we repeated our
calculations with different interactions and model spaces.  In $A=7$, the
effective operator is always a decided improvement but in $A=8$ and 10 the
results are more ambiguous.  Figure \ref{fig:cdb} and Tab.\ \ref{tab:cdb}
present results of calculations in a $\mathrm{N}_{\mathrm{max}}=8$ space with
the CD-Bonn interaction, conditioned as described in section \ref{s:methods}
for $A=8$ and 10. (We do not show $A=7$, and the size of the problem in
$^{10}$Be limits us to $\mathrm{N}_{\mathrm{max}}=6$ in $A=10$.) Once
again the effective operator appears to be an improvement in both cases, but
now, as Tab.\ \ref{tab:cdb} shows, the effective-operator curve for $A=8$
cancels itself \emph{too much} in the integral. And in $A=10$ the effective
operator does better than the figure indicates it should.

\begin{table}[t]
\caption{\label{tab:cdb} Matix elements $\mathcal{M}_{fi}$ produced by the
distributions $C(r)$ in Fig.~\ref{fig:cdb}.}
\begin{tabular}{lcc}
& 8 & 10 \\
\hline
full & -0.41 & -0.67 \\
bare & -0.48 & -0.80 \\
effective & -0.03 & -0.68 
\end{tabular}
\end{table}

\begin{figure}[b]
\includegraphics[width=\columnwidth]{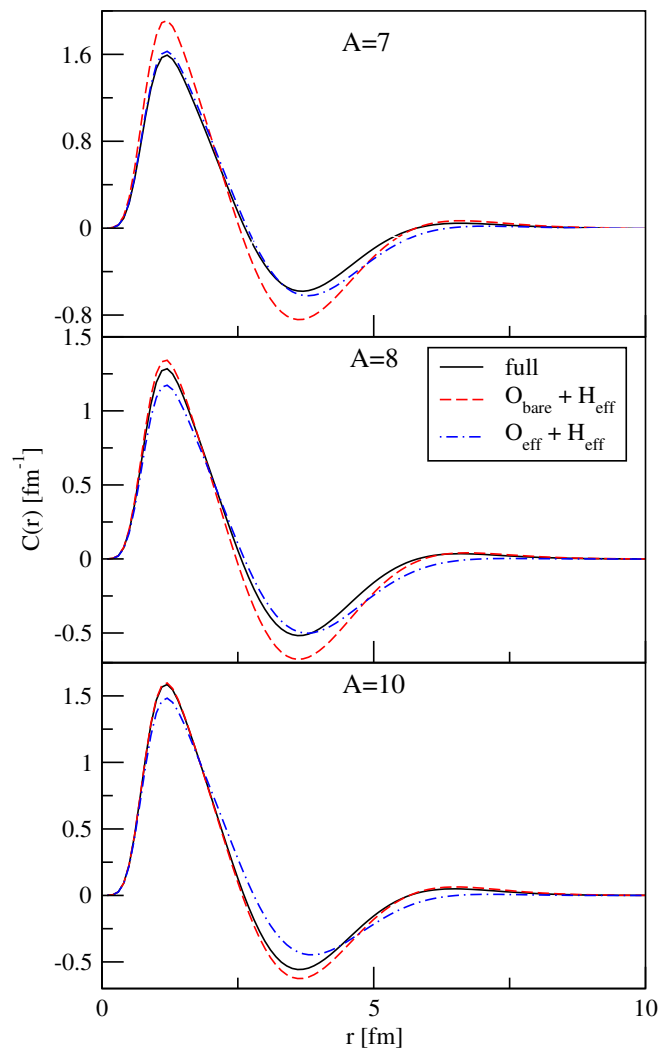}
\caption{\label{fig:srgn2} (Color online) Same as Fig.\ \ref{fig:srgn6} but
carried out in a small model space with $\mathrm{N}_{\mathrm{max}}=2$.}
\end{figure}

\begin{table}[b]
\caption{\label{tab:srgn2}Matix elements $\mathcal{M}_{fi}$ produced by the
distributions $C(r)$ in Fig. \ref{fig:srgn2}.}
\begin{tabular}{lccc}
& 7 & 8 & 10 \\
\hline
full & -1.06 & 0.70 & 1.10 \\
bare &  -0.90 & 0.37 & 0.99 \\
effective & -0.92 & 0.45 & 1.14
\end{tabular}
\end{table}

\begin{figure*} 
\includegraphics[width=32pc]{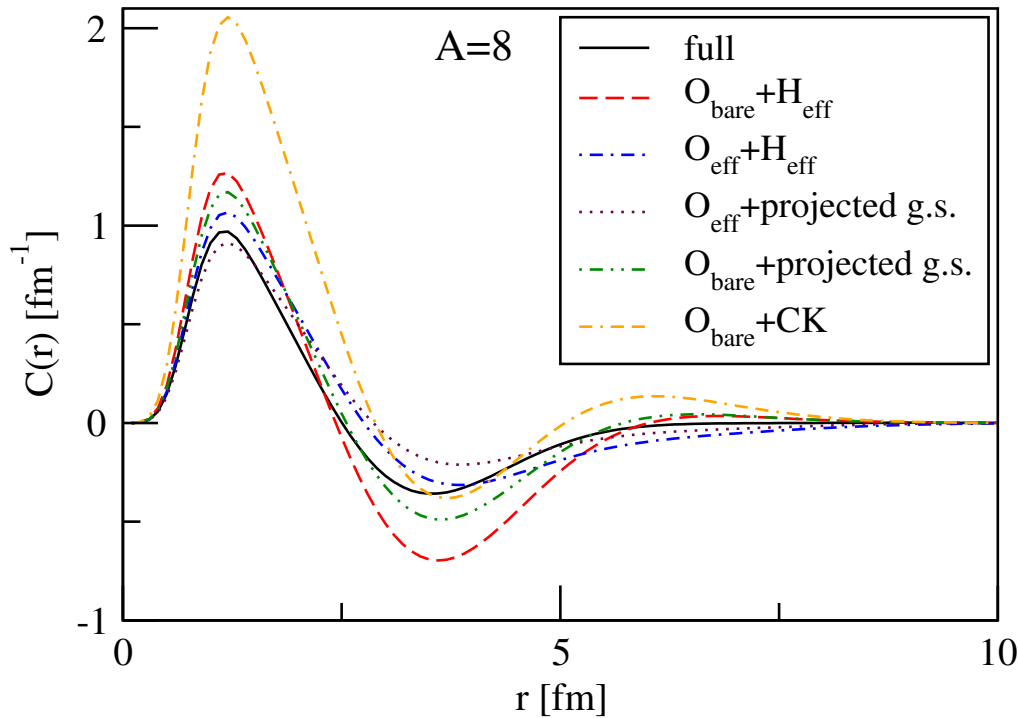}
\caption{\label{fig:a8srgn6-pgs} (Color online) $C(r)$ for $^8$He $\rightarrow
^8$Be: the solid (black), dashed (red) and dot-dashed (blue) lines are as in
Fig.\ \ref{fig:srgn6}.  The dotted (maroon) and the dot-dot-dashed (green)
curves result from using the normalized $p-$shell projections of the full
ground states with the effective and the bare decay operator respectively.  The
dot-dashed-dashed (orange) curve is the result with bare decay operator and the
Cohen-Kurath (CK) interaction \cite{coh65}.}
\end{figure*}

One might expect the procedure to work better when the full model space is
smaller, and/or when the full results differ less from the bare $p$-shell
prediction but neither of those are entirely the case either.  Figure
\ref{fig:srgn2} and Tab.~\ref{tab:srgn2} show results of calculations in which
we use the SRG-evolved interaction in a severely truncated space:
$\mathrm{N}_{\mathrm max}=2$.  Because we are using the same effective
interaction as before, we have reduced the amounts of the wave functions that
lie outside the $p$ shell.  Now all curves are close to each other everywhere,
but the effective operator, as expected, is an improvement in $A=7$ and 8.  In
$A=10$, however, the bare operator is nearly perfect and the effective operator
worse (though now, as Tab.\ \ref{tab:srgn2} shows, it is accidentally better in
the integral).





It seems, then, that except in this last example the effective decay operator
is a decided improvement, but also that the oscillations in $C(r)$ can
sometimes negate its advantages.  The natural way to do better is by adding
three-body terms to the effective interaction and decay operator.  We will not
do so in this preliminary study, but can look into whether the discrepancy at
the two-body level is due mainly to the defects in the decay operator or in the
interaction.  If we were to carry out the Lee-Suzuki procedure in $A=8$ and 10,
the $p$-shell ground states would resemble the normalized $p$-shell projections
of the full ground states.  We can therefore use these normalized projections
as proxies for the states that would be produced by the complete $A$-body
Lee-Suzuki $p$-shell Hamiltonian.  Figure \ref{fig:a8srgn6-pgs} shows the
resulting $C(r)$ with both the bare and two-body effective operators, alongside
the curves already displayed in Fig.\ \ref{fig:srgn6}.  The performance of the
bare operator improves noticeably, so that it is about as good as the two-body
effective operator in conjunction with the two-body effective Hamiltonian; its
integral is $\mathcal{M}_{fi}=0.63$.  And interestingly, performance gets worse
when the effective two-body operator is used instead
($\mathcal{M}_{fi}=0.74$).  The first result indicates that three-and-more-body
terms in the Hamiltonian affect the matrix element, and the second that such
terms in the decay operator do as well.  

Figure \ref{fig:a8srgn6-pgs} has one other curve, produced by the bare operator
in conjunction with the phenomenological Cohen-Kurath potential \cite{coh65}.
We can't really expect this empirical potential to reproduce the results of an
NCSM calculation in a nucleus that isn't in reality even stable, but it is
nevertheless interesting to see how different its results are. 

As mentioned in the introduction, the Lee-Suzuki mapping is only one of an
infinite number of possible mappings.  We tested for better-performing mappings
in $A=6$ by rotating the Lee-Suzuki $p$-shell basis vectors by arbitrary real
angles.  Though we didn't exhaustively explore all such possibilities, we found
it difficult to improve on the Lee-Suzuki results.  With the SRG interaction in
$\mathrm{N}_{\text{max}}=6$, rotation of the two $2+$ vectors by $15^\circ$
improved the effective operator $C(r)$'s marginally, but nowhere did we see a
dramatic improvement.  

\section{\label{s:discussion}Discussion}
  
We have shown that a nonperturbative renormalization of the effective decay
operator improves the shell-model's ability to reproduce \bbz matrix elements
from ab initio calculations.  What do these results imply for the heavier
nuclei we really care about?  There is some reason to hope that a two-body
effective operator will perform even better than it did here.  Both QRPA and
shell-model calculations in these nuclei show that $C(r)$ with the bare
operator nearly vanishes beyond about 3 fm; there are no oscillations of
consequence in those curves \cite{simkovic08}.  For that reason, the
performance of the effective operator may not be degraded by the cancellations
that play such a large role here.  In any event it is clearly worth determining
the effective two-body operator.

There are several possible routes to that end.  Ref.\ \cite{jansen11} reports
coupled-cluster calculations in $^{6}$He, the first in a nucleus with two
valence nucleons.  Coupled-cluster techniques scale well to intermediate-mass
nuclei and it should be possible to treat the nuclei $^{58}$Ni, $^{58}$Cu, and
$^{58}$Zn with two nucleons in the $f_{5/2}pg_{9/2}$ shell (along with the
simpler nuclei with one nucleon in that shell).  Such calculations will play
the same role as the NCSM calculations here, allowing the extraction of a
two-body effective Hamiltonian and decay operator, which can then be used in a
shell-model calculation of the decay of e.g.\ $^{76}$Ge.  If the Hamiltonian
proves inadequate, a better phenomenological one can be substituted, though at
the obvious cost of consistency.  Other techniques, including many-body
perturbation theory with soft renormalization-group-produced interactions
\cite{coraggio09} and the in-medium similarity renormalization group
\cite{tsukiyama10} offer a path to effective operators as well.

For a truly accurate calculation we probably will need three-body decay
operators.  All the methods mentioned above can produce these in principle, but
at least a few years of development are required.  In the meantime, we can
examine the question of whether two- and three-body decay operators will be
sufficient (or whether horribly complicated four-body operators will be
required) by extending the $p$-shell tests we report here.  Work in that
direction is underway.

\begin{acknowledgments}
We thank Ionel Stetcu and Alexander Lisetskiy for useful discussions/work.
J.E.\ gratefully acknowledges the support of the U.S.\ Department of Energy
through Contract No.\ DE-FG02-97ER41019. P.N.\ acknowledges support from the
NSERC grant No. 401945-2011.  Prepared in part by LLNL under Contract
DE-AC52-07NA27344.
\end{acknowledgments}

\end{document}